# A Transport-Friendly NIC for Multicore/Multiprocessor Systems

Wenji Wu, Matt Crawford, Phil DeMar
Fermilab, P.O. Box 500, Batavia, IL 60510

**Abstract**

Receive side scaling (RSS) is a network interface card (NIC) technology. It provides the benefits of parallel receive processing in multiprocessing environments. However, existing RSS-enabled NICs lack a critical data steering mechanism that would automatically steer incoming network data to the same core on which its application process resides. This absence causes inefficient cache usage if an application is not running on the core on which RSS has scheduled the received traffic to be processed. In Linux systems, it cannot even ensure that packets in a TCP flow are processed by a single core, even if the interrupts for the flow are pinned to a specific core. This results in degraded performance. In this paper, we develop such a data steering mechanism in the NIC for multicore or multiprocessor systems. This data steering mechanism is mainly targeted at TCP, but it can be extended to other transport layer protocols. We term a NIC with such a data steering mechanism "A Transport Friendly NIC" (A-TFN). Experimental results have proven the effectiveness of A-TFN in accelerating TCP/IP performance.

## 1. Introduction & Motivation

Computing is now shifting towards multiprocessing (e.g., SMT, CMP, SMP, and UNMA). The fundamental goal of multiprocessing is improved performance through the introduction of additional hardware threads, CPUs, or cores (all of which will be referred to as "cores" for simplicity). The emergence of multiprocessing has brought both opportunities and challenges for TCP/IP performance optimization in such environments. Modern network stacks can exploit parallel cores to allow either message-based parallelism or connection-based parallelism as a means of enhancing performance [1]. To date, major network stacks such as Windows, Solaris, Linux, and FreeBSD have been redesigned and parallelized to better utilize additional cores. While existing OSes exploit parallelism by allowing multiple threads to carry out network operations concurrently in the kernel, supporting this parallelism carries significant costs, particularly in the context of contention for shared resources, software synchronization, and poor cache efficiencies [2][3]. However, various optimization efforts, such as fine-grained locking and the read-copy-update technologies, have been helpful. While these optimizations definitely help improve TCP/IP processing in multiprocessing environments, they alone are not sufficient to keep pace with network speeds. A scalable, efficient network I/O in multiprocessing environments requires further optimization and coordination across all layers of the network stack, from network interface to application. Investigations regarding processor affinity [4][5][6][7] indicate that the coordinated affinity scheduling of protocol processing and network applications on the same target cores can significantly reduce contention for shared resources, minimize software synchronization overheads, and enhance cache efficiency.

Coordinated affinity scheduling of protocol processing and network applications on the same target cores has the following goals: (1) *Interrupt affinity,* network interrupts of the same type should be directed to a single core. Redistributing network interrupts in either a random or round-robin fashion to different cores has undesirable side effects [6]. (2) *Flow affinity,* packets of each flow should be processed by a single core. Flow affinity is especially important for TCP. TCP is a connection-oriented protocol, and it has a large and frequently accessed state that must be shared and protected when packets from the same connection are processed. Ensuring that all packets in a TCP flow are processed by a single core reduces contention for shared resources, minimizes software synchronization, and enhances cache efficiency. (3) *Network data affinity,* incoming network data should be steered to the same core on which its application process resides. This is becoming more important with the advent of Direct Cache Access (DCA) [8][9]. DCA is a NIC technology that seeks to directly place a received packet into a core's cache for immediate access by the protocol stack and application. Network data affinity maximizes cache efficiency and reduces core-to-core synchronization. In a multicore system, the function of network data steering is executed by directing the corresponding network interrupts to a specific core (or cores).

RSS [10] is a NIC technology. It supports multiple receive queues and integrates a hashing function in the NIC. The NIC computes a hash value for each incoming packet. Based on hash values, NIC assigns packets of the same data flow to a single queue and evenly distributes traffic flows across queues. With Message Signal Interrupt (MSI/MSI-X) [11] support, each receive queue is assigned a dedicated interrupt and

RSS steers interrupts on a per-queue basis. RSS provides the benefits of parallel receive processing in multiprocessing environments. Operating systems like Windows, Solaris, Linux, and FreeBSD now support interrupt affinity. When an RSS receive queue (or interrupt) is tied to a specific core, packets from the same flow are steered to that core (Flow pinning [12]). This ensures flow affinity on most OSes, except for Linux (see Section 2).

However, RSS has a limitation: it cannot steer incoming network data to the same core where its application process resides. The reason is simple: the existing RSS-enabled NICs do not maintain the relationship "Traffic Flows → Network applications → Cores" in the NIC (since network applications run on cores, the most critical relationship is simply "Traffic Flows → Cores (Applications)") and the existing OSes do not support such capability. This is symptomatic of a broader disconnect between existing software architecture and multicore hardware. On OSes like Windows, if an application is not running on the core on which RSS has scheduled the received traffic to be processed, network data affinity cannot be achieved, resulting in degraded cache efficiency [10]. This limitation might cause serious performance degradation for NUMA systems. Furthermore, On OSes like Linux, if an application runs on cores other than those where its corresponding RSS network interrupts are directed, Linux TCP processing might alternate between different cores even if the interrupts for the flow are pinned to one core [see Section 2]. There will be neither flow affinity, nor network data affinity. As a result, it will lead to poor cache efficiency and cause significant core-to-core synchronization overheads. The overall system efficiency could be severely degraded.

The NIC technologies, such as Intel's VMDq [27] or the PCI-SIG's SR-IOV [28], do provide data steering capabilities for the NICs. But they are I/O virtualization technologies targeting at virtual machines in the virtualized environment, with different research issues.

In this paper, we propose a NIC mechanism to remedy the RSS limitation. It steers incoming network data to the same core on which its application resides. Our data steering mechanism is mainly targeted at TCP, but can be extended to UDP and SCTP. We term a NIC with such a data steering mechanism A Transport-Friendly NIC, or A-TFN. The basic idea is simple: A-TFN maintains the relationship "Traffic Flows → Cores (Applications)" in the NIC, and OSes are correspondingly enhanced to support such capability. For transport layer traffic, A-TFN maintains a Flow-to-Core table in the NIC, with one entry per flow; each entry tracks which receive queue (core) a flow should be assigned to. A-TFN makes use of the facts that (1) TCP connections always involve packets flowing in both directions (ACKs, if nothing else). And (2) when an application makes socket-related system calls, that calling application's context may be borrowed to carry out network processing in process context. With each outgoing transport-layer packet, the OS records a processor core ID and uses it to update the entry in the Flow-to-Core table. As soon as any network processing is performed in a process context, A-TFN learns of the core on which an application process resides and can steer future incoming traffic to the right core. Clearly, to design such a mechanism, there is an obvious trade-off between the amount of work done in the NIC and in the OS. In the paper, we discuss two design options. Option 1 is to minimize changes in the OS and focuses instead on identifying the minimal set of mechanisms to add to the NIC. Clearly, this design adds complexity and cost to the NIC. On the other end of the design space, it could be let the OS update the flow-to-core table directly without changing anything in the NIC hardware (option 2). Conceptually, this approach could be fairly straightforward to implement. However, it might add significant extra communication overheads between the OS and the NIC, especially when the Flow-to-Core table gets large. Due to space limitation, this paper is mainly focused on the first design option. The new NIC is emulated in software and it shows that the solution is effective and practical to remedy RSS's limitation. In our future work, we will explore the second design option. The contributions of this paper are threefold. First, we show for certain OSes, such as Linux, that tying a traffic flow to a single core does not necessarily ensure flow affinity or network data affinity. Second, we show that existing RSS-enabled NICs lack a mechanism to automatically steer packets of a data flow to the same core(s), where they will be protocol-processed and finally consumed by the application. This is symptomatic of a broader disconnect between existing software architecture and multicore hardware. Third, we develop such a data steering mechanism in the NIC for multi-core or multiprocessor systems.

The remainder of the paper is organized as follows: In Section 2, we present problem formulation. Section 3 describes the A-TFN mechanism. In section 4, we discuss experiment results that showcase the effectiveness of our A-TFN mechanism. In section 5, we present related research. We conclude in section 6.

## 2. Problem Formulation

### 2.1 Packet Receive-Processing with RSS

RSS is a NIC technology. It supports multiple receive queues and integrates a hashing function in the

NIC. NIC computes a hash value for each incoming packet. Based on hash values and an indirection table, NIC assigns packets of the same data flow to a single queue and evenly distributes traffic flows across queues. With Message Signal Interrupt (MSI/MSI-X) and Flow Pinning support, each receive queue is assigned a dedicated interrupt and tied to a specific core. The device driver allocates and maintains a ring buffer for each receive queue within system memory. For packet reception, a ring buffer must be initialized and pre-allocated with empty packet buffers that have been memory-mapped into the address space that is accessible by the NIC over the system I/O bus. The ring buffer size is device- and driver-dependent. Fig. 1 illustrates packet receive-processing with RSS: (1) When incoming packets arrive, the hash function (e.g., Toeplitz hashing [10]) is applied to the header to produce a hash result. The hash type, which is configurable, controls which incoming packet fields are used to generate the hash result. OSes can enable any combination of the following fields: *source address, source port, destination address, destination port,* and *protocol*. The hash mask is applied to the hash result to identify the number of bits that are used to index the indirection table. The indirection table is the data structure that contains an array of core numbers to be used for RSS. Each lookup from the indirection table identifies the core and hence, the associated receive queue. (2) The NIC assigns incoming packets to the corresponding receive queues. (3) The NIC DMAs (direct memory access) the received packets into the corresponding ring buffers in the host system memory. (4) The NIC sends interrupts to the cores that are associated with the non-empty queues. Subsequently, the cores respond to the network interrupts and process received packets up through the network stack from the corresponding ring buffers one by one.

The OS can periodically rebalance the network load on cores by updating the indirection table, based on the assumption that the hash function will evenly distribute incoming traffic flows across the indirection table entries. Since the OS does not know which specific entry in the indirection table an incoming traffic flow will be mapped to, it can only passively react to load imbalance situations by changing each core's number of appearances in the indirection table. For better load balancing performance, the size of the indirection table is typically two to eight times the number of cores in the system [10]. For example, in Fig. 1, the indirection table has 8 entries, which are populated as shown. As such, traffic loads directed to Core 0, 1, 2, and 3 are 50%, 25%, 12.5%, and 12.5%, respectively. Some OSes like Linux and FreeBSD do not support the function of an indirection table; the incoming packets are directly mapped to the receive queues. These OSes cannot perform dynamic load balancing.

**2.2 RSS Limitation and the Reasons**

RSS provides the benefits of parallel receive processing. However, this mechanism does present certain limitation: it cannot steer incoming network data to the same core on which its application resides.

The reason is simple: the existing RSS-enabled NICs do not maintain the relationship "Traffic Flows → Network applications → Cores" in the NIC (since network applications run on cores, the most critical relationship is simply "Traffic Flows → Cores (Applications).") and the existing OSes do not support such capability. When packets arrive, the hash function is applied to the header to produce a hash result. Based on the hash values, the NIC assigns packets to receive queues and then cores, with no way to consider on which core the corresponding application is running. Although receive queues can be instructed to send interrupt to a specific set of cores, existing general purpose OSes can only provide limited process-to-interrupt affinity capability; network interrupt delivery is not synchronized with process scheduling. This is because the OS schedulers have other priorities, such as load balancing and fairness, over process-to-interrupt affinity. Besides, multiple network applications' traffic might map to a single interrupt, which brings new challenges to an OS scheduler. Therefore, a network application might be scheduled on cores other than those where its corresponding network interrupts are directed. This is symptomatic of a broader disconnect between existing software architecture and multicore hardware

OSes like Windows implement the function of the indirection table, which can provide limited data steering capabilities for RSS-enabled NICs. However, it still cannot steer packets of a data flow to the same core where the application process resides. Turning again to

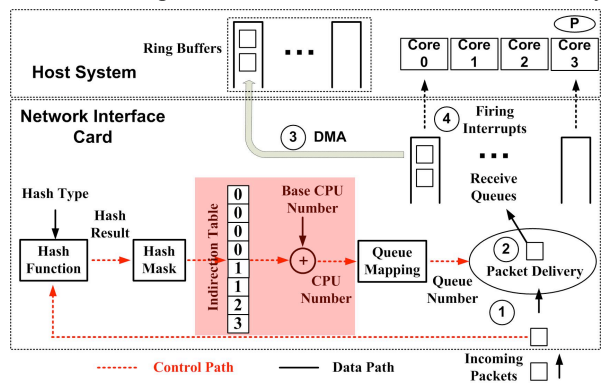

**Fig. 1 Packet Receiving Process with RSS**

Fig 1, process P is scheduled to run on Core 3. Its traffic might be hashed to an entry that directs to other cores. The OS does not know which specific entry in the indirection table a traffic flow will be mapped to.

With existing RSS capability, there are many cases in OSes in which a network application resides on cores other than those to which its corresponding network interrupts are directed: (1) A single-threaded application might handle multiple concurrent TCP connections. Assuming such an application handles $n$ concurrent TCP connections and runs on an $m$-core system, an RSS-enabled NIC will evenly (statistically) distribute the $n$ connections across the $m$ cores. Since the application can only run on a single core at any moment, only $n/m$ connections' network interrupts are directed to the same core where the application runs. (2) Soft partition technologies like CPUSET [13] are applied in the context of networking environments. Since the OS (or system administrator) has no way of knowing to which specific core they will be mapped, network applications might be soft-partitioned on cores other than those to which their network interrupts are directed. (3) The general purpose OSes scheduler prioritizes load balancing or power saving over process-to-interrupt affinity [14][15]. For OSes like Linux, when the multicore peak performance mode is enabled, the scheduler tries to use all cores in parallel to the greatest extent possible, distributing the load equally among them. When the multicore power saving mode is enabled, the scheduler is biased to restrict the workload to a single physical processor. As a result, a network application might be scheduled on cores other than those to which its network interrupts are directed.

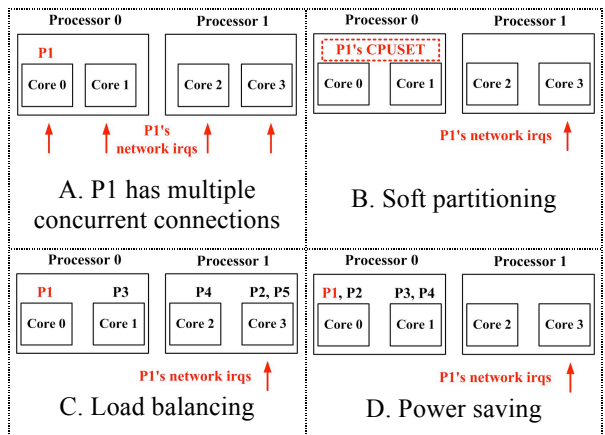

**Fig. 2 Network Irqs and Apps. on Different Cores**

For clarity, we illustrate the above cases in Fig. 2. The system contains two physical processors, each with two cores. P1 – P5 are processes that run within the system. P1 is a network process that includes traffic flows. An RSS-enabled NIC steers the traffic flows to different cores, as shown in the figure (red arrows). In all of these cases, P1 resides on cores other than those to which its corresponding network interrupts are directed.

On OSes like Windows, when a core responds to the network interrupt, the corresponding interrupt handler is called, within which a deferred procedure call (DPC) is scheduled. On the core, DPC processes received packets up through the network stack from the corresponding ring buffer one by one [16]. Therefore, on Windows, tying a traffic flow to a single core does ensure interrupt affinity and flow affinity. However, if network interrupts are not directed to cores on which the corresponding applications reside, network data affinity cannot be achieved, resulting in degraded cache efficiency [10]. This reality might cause serious performance degradation for NUMA systems.

On some OSes, like Linux, tying a traffic flow to a single core does not necessarily ensure flow affinity or network data affinity due to Linux TCP's unique prequeue-backlog queue design. In the following sections, we discuss in detail why the combination of RSS and Flow Pinning cannot ensure flow affinity and network data affinity in Linux.

## 2.3 Linux Network Processing in Multicore Systems

As a modern parallel network stack, Linux exploits packet-based parallelism, which allows multiple threads to simultaneously process different packets from the same or different connections. Two types of threads may perform network processing in Linux: application threads in process context and interrupt threads in interrupt context. When an application makes socket-related system calls, that application's process context may be borrowed to carry out network processing. When a NIC interrupts a core, the associated handler services the NIC and schedules the softirq, softnet. Afterwards, the softnet handler processes received packets up through the network stack in interrupt context. TCP is a connection-oriented protocol, and it has a large and frequently accessed state that must be shared and protected. In the case of the Linux TCP, the data structure socket maintains a connection's various TCP states, and there is a per-socket lock to protect it from unsynchronized access. The lock consists of a spinlock and a binary semaphore. The binary semaphore construction is based on the spinlock. In Linux, since an interrupt thread cannot sleep, when it accesses a socket, the socket is protected with the spinlock. When an application thread accesses a socket, the socket is locked with the binary semaphore and is considered "owned-by-user." The binary semaphore synchronizes multiple application threads among themselves. It is also used as a flag to notify interrupt threads that a socket is "owned-by-user" to coordinate synchronized access to the socket between interrupt and application threads.

Our previous research [17][18] studied the details of the Linux packet receiving process. Here, we simply summarize Linux TCP processing of the data receive path in interrupt and process contexts, respectively.

*a) TCP Processing in Interrupt Context*
(1) When the NIC interrupts a core, the network interrupt's associated handler services the NIC and schedules the softirq, softnet.
(2) The softnet handler moves a packet from the ring buffer and processes the packet up through the network stack. If there is no packet available in the ring buffer, the softnet handler exits.
(3) A TCP packet (segment) is delivered up to the TCP layer. The network stack first tries to identify the socket to which the packet belongs, and then seeks to lock (spinlock) the socket.
(4) The network stack checks if the socket is "owned-by-user" or if an application thread is sleeping and awaiting data:
    - If yes, the packet will be enqueued into the socket's backlog queue or prequeue. TCP processing will be performed later in process context by the application thread.
    - If not, the network stack will perform TCP processing on the packet in interrupt context.
(5) Unlock the socket; go to step 2.

*b) TCP Processing in Process Context*
(1) An application thread makes a socket-related receive system call.
(2) Once the system call reaches the TCP layer, the network stack seeks to lock (semaphore) the socket first.
(3) The network stack moves data from the socket into the user space.
(4) If the socket's prequeue and/or backlog queue are not empty, the calling application's process context would be borrowed to carry out TCP processing.
(5) Unlock the socket and return from the system call.

For the data transmit path, network processing starts in the process context when an application makes socket-related system calls to send data. If TCP gives permission to send (based on TCP receiver window, congestion window, and sender window statuses), network processing in process context can reach down to the bottom of the protocol stack; otherwise, transmit side network processing is triggered by incoming TCP ACKs for the data receive path, which are performed in their execution environments (interrupt or process contexts). In this paper, we focus mainly on receive side processing because it is known to be more memory intensive and complex. Furthermore, TCP processing on the transmit side is also dependent on ACKs in the data receive path.

As described above, whether TCP processing is performed in process or interrupt contexts depends on the volatile runtime environments. For example, we used FTP to download Linux kernels from www.kernel.org; we instrument the Linux network stack to record the percentage of traffic processed in process context. The recorded percentage ranges from 50% to 75%.

It is clear that, in a multicore system, when an application's process context is borrowed to execute the network stack, TCP processing is performed on the core(s) where the application is scheduled to run. When TCP processing is performed in interrupt context, it is performed on the cores to which the network interrupts are directed. Take, for example, Fig. 3, in which network interrupts are directed to core 0 and the associated network application is scheduled to run on core 1. In interrupt context, TCP is processed on core 0; in process context, this occurs on core 1. Since TCP processing performed in process or interrupt contexts depends on volatile runtime conditions, it may alternate between these two cores. Therefore, although the combination of RSS and Flow Pinning can tie a traffic flow to a single core, when a network application resides on some other core, TCP processing might alternate between different cores. We would achieve neither flow affinity nor network data affinity.

**2.4 Negative Impacts**

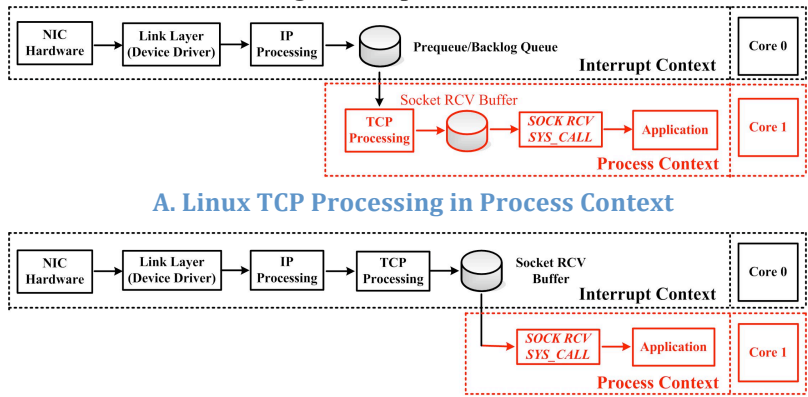

A. Linux TCP Processing in Process Context

B. Linux TCP Processing in Interrupt Context

Fig. 3 Linux TCP Processing Contexts in the Data Receive Path

If an application runs on cores other than those where its corresponding RSS network interrupts are directed, various negative impacts result. On both Windows and Linux systems, network data affinity cannot be achieved. Furthermore, on OSes like Linux, TCP processing might alternate between different cores even if the interrupts for the flow are pinned to a specific core. As a result, it will lead to poor cache efficiency and cause significant core-to-core synchronization overheads. Also, it renders the DCA technology ineffective. In multiple core systems, core-to-core synchronizations involve costly snoops and MESI operations [19], resulting in extra system bus traffic. This is especially expensive when the contending cores exist within different physical processors, which usually involves synchronous read/write operations to a certain memory location.

In addition, for Linux, interrupt and application threads contend for shared resources, such as locks, when they concurrently process packets from the same flow. The socket's spinlock, for example, would be in severe contention. When a lock is in contention, contending threads simply wait in a loop ("spin"), repeatedly checking until the lock becomes available. While waiting, no useful work is executed. Contention for other shared resources, such as memory and system bus, also occurs frequently. Since this intra-flow contention may occur on a per-packet basis, the total contention overhead could be severe in high bandwidth network environments.

To demonstrate the negative impacts, we ran data transmission experiments over an isolated sub-network. The sender and receiver's detailed features are as:

**Sender:** Dell R-805. CPU: two Quad Core AMD Opteron 2346HE, 1.8GHz, HT1. NIC: Broadcom NetXtreme II 5708, 1Gbps, DCA not supported. OS: Linux 2.6.28.

**Receiver:** SuperMicro Server. CPU: two Intel Xeon CPUs, 2.66 GHz, Family 6, Model 15. NIC: Intel PRO/1000, 1Gbp, DCA not supported. OS: Linux 2.6.28. The receiver's CPU architecture is as shown in Fig. 4.

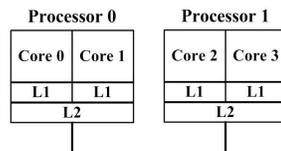

**Fig. 4 Receiver CPUs**

In the experiments, we used *iperf* [20] to send data in one direction. The sender transmitted one TCP stream to the receiver for 100 seconds. In the receiver, network interrupts were all directed to core 0. However, iperf was pinned to different cores: (1) Iperf was pinned to core 0 (network interrupts and applications were pinned to the same core). (2) Iperf was pinned to core 1 (network interrupts and applications were pinned to different cores, but within the same processor). (3) Iperf was pinned to core 2 (network interrupts and applications were pinned to different processors). The throughput rates in these experiments all saturated the 1Gbps link (around 940 Mbps). The experiments were designed to feature the same throughput rates for the sake of better comparisons.

We ran *oprofile* [21] to profile system performance in the case of the receiver. The metrics of interest were: INST_RETIRED, the number of instructions retired; BUS_TRAN_ANY, the total number of completed bus transactions; and BUS_HITM_DRV, the number of HITM (hit modified cache line) signals asserted [22]. For these metrics, the number of events between samples was 10000. We also enabled the Linux Lockstat [13] to collect lock statistics. On this basis we calculated the total time spent waiting to acquire various kernel locks, and we called this WAITTIME-TOTAL. Consistent results were obtained across repeated runs. The results are as listed in Fig. 5, with a 95% confidence interval.

The throughput rates in these experiments all saturated the 1Gbps link. However, Fig. 5 clearly shows that the metrics of iperf @ Core 1 and Core 2 are much higher than those of iperf @ Core 0. This clearly verifies that when a network application is scheduled on cores other than those to which the corresponding network interrupts are directed, severely degraded system efficiency will result.

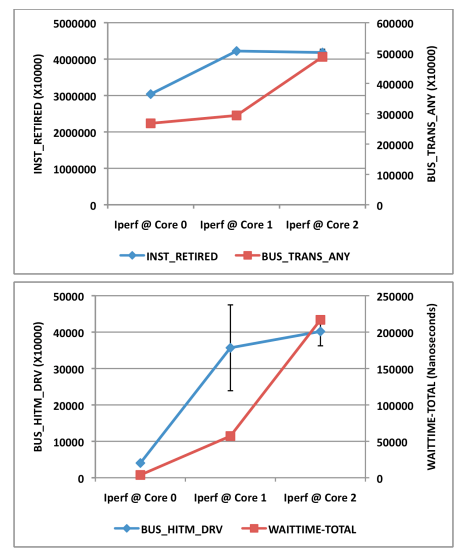

**Fig. 5. Experiment Results**

INST_RETIRED measures the load on the receiver. The results clearly show that contention for shared resources between interrupt and application threads led to an extra load. The extra load is mainly related to time spent waiting for locks. The experimental WAITTIME-TOTAL data verify this point. It is surprising that the BUS_TRANS_ANY of iperf @ Core 2 is almost twice that of iperf @ Core 0. The BUS_HITM_DRV of iperf @ Core 0 is far less that that of iperf @ Core 1 and Core 2. Since the throughput rates in these experiments all saturated the 1Gbps link, the extra BUS_TRANS_ANY and BUS_HITM_DRV transactions of iperf @ Core 1 and Core 2 were caused by cache trashing and lock contention, as analyzed above.

## 3. A Transport Friendly NIC (A-TFN)

### 3.1 A-TFN Design Principles & Alternatives

Previous analyses and experiments clearly show that, existing RSS-enabled NICs cannot automatically steer incoming network data to the core on which its application process resides. In this paper, we propose a NIC mechanism to remedy this limitation. It steer packets of a data flow to the same cores where they will be protocol-processed and subsequently consumed by the application. Our data steering mechanism is mainly targeted at TCP, but can be extended to UDP and SCTP. We term a NIC with such a data steering mechanism A Transport-Friendly NIC, or A-TFN.

A-TFN's basic idea is simple: it maintains the relationship "Traffic Flows → Cores (Applications) in the NIC and OSes are correspondingly enhanced to support such capability. For transport layer traffic, A-TFN maintains a Flow-to-Core table in the NIC, with one entry per flow; each entry tracks which receive queue (core) a flow should be assigned to. A-TFN makes use of the facts that (1) a TCP connection's traffic is bidirectional. For a unidirectional data flow, ACKs on the reverse path result in bidirectional traffic. And (2) when an application makes socket-related system calls, that application's process context might be borrowed to carry out network processing in process context. This is true for almost all OSes in the data transmit path. With each outgoing transport-layer packet, the OS records a processor core ID and uses it to update the entry in the Flow-to-Core table. As soon as any network processing is performed in a process context, A-TFN learns of the core on which an application process resides and can steer future incoming traffic to the right core.

Clearly, the design of such a mechanism involves a trade-off between the amount of work done in the NIC and in the OS. There are two design options. Option 1 is to minimize changes in the OS and focuses instead on identifying the minimal set of mechanisms to add to the NIC. Clearly, this design adds complexity and cost to the NIC. On the other end of the design space, it could be let the OS update the flow-to-core table directly without changing anything in the NIC hardware (option 2). Conceptually, this approach could be fairly straightforward to implement. However, it might add significant extra communication overheads between the OS and the NIC, especially when the Flow-to-Core table gets large. Due to space limitation, this paper is mainly focused on the first design option. In our future work, we will explore the second design option. Besides, option 1 design has other goals: (1) A-TFN must be simple and efficient. This is because the NIC controllers usually utilize a less powerful CPU with a simplified instruction set and insufficient memory to hold complex firmware. (2) A-TFN must preserve in-order packet delivery. (3) The communication overheads between the OS and A-TFN must be minimal.

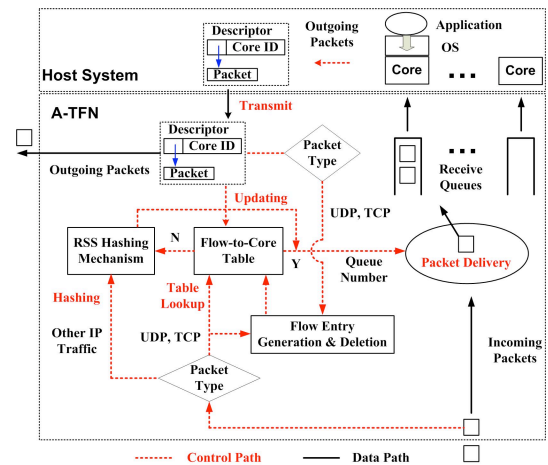

**Fig. 6 A-TFN Mechanisms**

### 3.2 A-TFN Details

Fig. 6 illustrates the A-TFN details. A-TFN extends the current RSS technologies. It supports multiple receive queues in the NIC, up to the number of cores in the system. With MSI and Flow-Pinning support, each receive queue has a dedicated interrupt and is tied to a specific core; each core in the system is assigned a specific receive queue. A-TFN handles non-transport layer traffic in the same way as does RSS. That is, based on a hash of the incoming packet's headers, the NIC assigns it to the same queue as other packets from the same data flow, and distributes different flows across queues. For transport layer traffic, A-TFN maintains a Flow-to-Core table with a single entry per flow. Each entry tracks the receive queue (core) to which a flow should be assigned. The entries within the Flow-to-Core table are updated by outgoing packets. For unidirectional TCP data flows, outgoing ACKs update the Flow-to-Core table. For an outgoing transport-layer packet, the OS records a processing core ID in the transmit descriptor and passes it to the NIC. Since each packet contains a complete identification of the flow it belongs to, the specific Flow → Core relationship could be effectively extracted from the outgoing packet and its accompanying transmit descriptor. As soon as any network processing is performed in the process context, A-TFN learns of on which core an application resides.

### 3.3 Flow-to-Core Table and its Operations

The Flow-to-Core table appears in Fig. 7. Flow entries are managed in a hash table, with a linked list to resolve collisions. Each entry consists of:

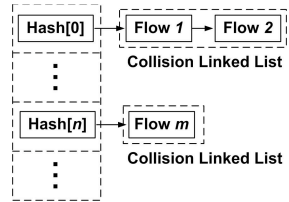

Fig. 7 Flow-to-Core Table

- **Traffic Flow**. A-TFN makes use of the 5-tuple *{src_addr, dst_addr, protocol, src_port, dst_port}* in the receive direction to specify a flow. Therefore, for an outgoing packet with the header *{(src_addr: x), (dst_addr: y), (protocol: z), (src_port: p), (dst_port: q)}*, its corresponding flow entry in the table is identified as *{(src_addr: y), (dst_addr: x), (protocol: z), (src_port: q), (dst_port: p)}*.
- **Core ID**. The core to which the flow should be steered.
- **Transition State**. A flag to indicate if the flow is in a transition state. The goal is to ensure in-order packet delivery.
- **Packets in Transition**. A simple packet list to accommodate temporary packets when the flow is in a transition state. The goal is to ensure in-order packet delivery.

In addition, to avoid non-deterministic packet processing time, a collision-resolving linked list is limited to a maximum size of *MaxListSize*. Flows are not evicted in case of collision. When a specific hash's collision-resolving list reaches *MaxListSize*, later flows with that hash will not enter into the table.

*a). Flow Entry Generation and Deletion*

A-TFN monitors each incoming and outgoing packet to maintain the Flow-to-Core Table. An entry is generated in the Flow-to-Core table as soon as A-TFN detects a successful three-way handshake. However, to reduce NIC complexity, A-TFN need not run a full TCP state machine in the NIC. A flow entry is deleted after a configurable period of time, $T_{delete}$, has elapsed without traffic. In this way, A-TFN need not handle all exceptions such as missing FIN packets and various timeouts.

To prevent memory exhaustion or malicious attacks, A-TFN sets an upper bound on the number of entries in the Flow-to-Core Table. When the Flow-to-Core table starts to become full, TCP flows can be aged out more aggressively by using a smaller $T_{delete}$. For traffic flows that are not in the Flow-to-Core table, packets are delivered based on a hash of the incoming packets' headers.

*b). Flow Entry Updating*

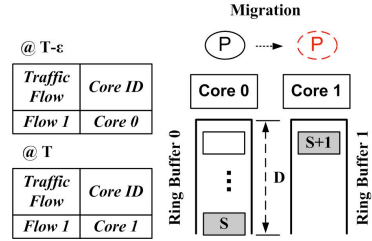

Fig. 8 A Simplified Model for Packet Reordering Analysis

The entries of the Flow-to-Core table are updated by outgoing packets. For each outgoing transport-layer packet, the OS records a processing core ID in the transmit descriptor and passes it to the NIC. A naive way to update the corresponding flow entry is with the passed core ID, omitting any other measures. As soon as any network processing is performed in process context, A-TFN learns of the process migration and can steer future incoming traffic to the right core. However, this simple flow entry updating mechanism cannot guarantee in-order packet delivery. TCP performance suffers in the event of severe packet reordering [23]. In the following sections, we use a simplified model to analyze why this approach cannot guarantee in-order packet delivery.

As shown in Fig. 8, at time $T-\varepsilon$, *Flow 1*'s flow entry maps to Core 0 in the Flow-to-Core table. At this instant, packet *S* of *Flow 1* arrives; based on the Flow-to-Core table, it is assigned to Core 0. At time $T$, due to process migration, *Flow 1*'s flow entry is updated and maps to Core 1. At $T+\varepsilon$, Packet *S+1* of *Flow 1* arrives and is assigned to the new core, namely Core 1. As described above, after assigning received packets to the corresponding receive queues, A-TFN copies them into system memory via DMA, and finally fires network interrupts if necessary. When a core responds to a network interrupt, it processes received packets up through the network stack from the corresponding ring buffer one by one. In our case, Core 0 processes packet S up through the network stack from Ring Buffer 0, and Core 1 services packet S+1 from Ring Buffer 1. Let $T_{service}(S)$ and $T_{service}(S+1)$ be the times at which the network stack starts to service packets S and S+1, respectively. If $T_{service}(S) > T_{service}(S+1)$, the network stack would see packet S+1 earlier than packet S, resulting in packet reordering. Let D be the ring buffer size and let the network stack's packet service rate be $R_{service}$ (packets per second). Assume there are $n$ packets ahead of S in Ring Buffer 0 and $m$ packets ahead of S+1 in Ring Buffer 1. Then:

$$T_{service}(S) = T - \varepsilon + n/R_{service} \quad (1)$$
$$T_{service}(S+1) = T + \varepsilon + m/R_{service} \quad (2)$$

If $\varepsilon$ is small, the condition of $T_{service}(S) > T_{service}(S+1)$ would hold and lead to packet reordering. Since the ring buffer size is $D$, the worst case is $n = D-1$ and $m = 0$:

$$T_{service}(S) = T - \varepsilon + (D-1)/R_{service} \quad (3)$$
$$T_{service}(S+1) = T + \varepsilon \quad (4)$$

However, if the delivery of packet S+1 to Core 1 can be delayed for at least $(D-1)/R_{service}$, then $T_{service}(S+1) \geq T + \varepsilon + (D-1)/R_{service}$. As a result, $T_{service}(S+1) > T_{service}(S)$ and in-order packet delivery can be guaranteed.

Therefore, A-TFN adopts the following flow entry updating mechanism: for each outgoing transport-layer packet, the OS records a processing core ID in the transmit descriptor and passes it to the NIC to update the corresponding flow entry. For a TCP flow entry, if the new core id is different from the old one, the flow enters the "transition" state. Correspondingly, its "***Transition State***" is set to "**Yes**" and a timer is started for this entry. The timer's expiration value is set to $T_{timer} = (D-1)/R_{service}$. Incoming packets of a flow in the transition state are added to the tail of "***Packets in Transition***" instead of being immediately delivered. When the timer expires, the flow leaves the transition state. The "***Transition State***" is set back to "**No**" and all of the packets in "***Packets in Transition,***" if they exist, are assigned to the new core. For a flow in the "non-transition" state, its packets are directly steered to the corresponding core. The ring buffer size D is a design parameter for the NIC and driver. For example, the Myricom 10Gb NIC is 512, and Intel's 1Gb NIC is 256. With current computing power, $(D-1)/R_{service}$ is usually at the sub-millisecond level, at best. For A-TFN, $T_{timer}$ is a design parameter and is configurable.

### 3.4 Required OS Support

Option 1's A-TFN design requires only two small OS changes in order to be properly supported. These can be easily implemented. (1) For an outgoing transport-layer packet, the OS needs to record a processing core ID in the transmit descriptor passed to the NIC. (2) The transmit descriptor needs to be updated with a new element to store this core ID. A single-byte element can support up to 256 cores, which is sufficient for most of today's systems. In addition, the size of a transmit descriptor is usually small, typically less than a cache line. Transmit descriptors are usually copied to the NIC by DMA using whole cache line memory transactions. Adding a byte to the transmit descriptor introduces almost no extra communication overhead between the OS and NIC.

### 4. Analysis and Experiments

The A-TFN mechanism is simple. It guarantees in-order packet delivery and requires the most minimal OS support. In addition, the communication overheads between the OS and A-TFN are reduced to a minimum. Compared to the extensively pursued TOE (TCP Offloading Engine) technology, which seeks to offload processing of the engine TCP/IP stack to the NIC, A-TFN is much less complex: (1) A-TFN does not require a complicated TCP engine within the NIC; (2) there is no need to synchronize TCP flow states between the OS and A-TFN; and (3) there is no need to enforce flow access control in the NIC. Therefore, with the latest hardware and software technologies, A-TFN can be effectively implemented. In the following sections, we use a combination of analytical and experimental techniques to evaluate the effectiveness of A-TFN mechanisms.

### 4.1 Analytical Evaluation

*a) Delay*

To ensure in-order packet delivery, incoming packets of a flow in the transition state are added to the tail of "***Packets in Transition***". These packets are delivered later, when the flow exits the transition state. Obviously, this mechanism not only adds delay to certain packets but also requires extra memory to accommodate them. However, considering that the scheduler of a general purpose OS tries to the extent possible to schedule a process onto the same core as that on which it was previously running, process migration does not occur frequently. As a result, only a very few packets will be held and will experience extra delay. Our experiments in Section 4.2 confirm this point. Clearly, the maximum delay a held packet can experience is $T_{timer}$. Previous analysis has shown that in-order packet delivery is guaranteed when $T_{timer}$ is set to $(D-1)/R_{service}$. However, in the real world, incoming packets rarely fill a ring buffer. If $T_{timer}$ were configured to be smaller, this would still ensure in-order packet delivery in most cases. In [23], we record the duration for which the OS processes the ring buffer. Our experiments have clearly shown that the duration is generally shorter than 20 microseconds. In most cases the extra delay is so small that it can be ignored.

*b) Flow Affinity and Network Data Affinity*

The intent of A-TFN is to automatically steer incoming network data to the same core on which its application process resides. As soon as any network processing is performed in a process context, A-TFN learns of the core on which an application process resides and can steer future incoming traffic to the right core. As a result, the desired flow affinity and network data affinity are guaranteed. For Linux, in rare circumstances, network processing might always occur in the interrupt context. When this happens, A-TFN cannot learn of process migration and cannot steer incoming traffic to the cores on which the corresponding applications are running. Flow affinity still holds, but, network data affinity does not in these cases.

*c) Hardware design considerations*

A-TFN's memory is mainly used to maintain the Flow-to-Core table, holding flow entries and accommodating packets for flows in the transition state. To hold a single flow entry, 20 bytes is quite sufficient. Therefore, a 10,000-entry Flow-to-Core table requires only 0.2 MB of memory. (These figures apply to IPv4; IPv6 support would add 24 bytes to the size of each entry, or less if the flow label could be relied upon.) In addition, to accommodate packets for flows in transition, if $T_{timer}$ is set to 0.2 millisecond, even for a 10Gbps NIC, the memory required is *0.2 millisecond × 10Gbps = 0.25 MB,* at maximum. In the worst case, an extra 0.5 MB of fast SRAM is enough to support the Flow-to-Core Table. A Cypress 4Mb (10ns) SRAM now costs around $7, with ICC=90 mA@10ns and ISB2=10mA. Table 1 lists the cost, memory size and power consumption of three main 10G Ethernet NICs in the market. Clearly, A-TFN's requirement of an extra 0.5 MB fast SRAM in the NIC won't add much extra cost (< 1%) and power consumption (<5% for Intel and Chelsio; <10% for Myricom) to current 10Gbps NICs.

A linked list in HW is expensive to build given all the extra handling. However, there will be a tradeoff in hardware complexity and A-TFN effectiveness. We further discuss this later.

| Vendor | Cost | Memory | Power |
|---|---|---|---|
| Intel | $1500 | N/A | 10.4W |
| Chelsio | > $750 | 256MB | 16W |
| Myricom | $ 850 | 2MB SRAM | 4.1W |

**Table 1 10G PCI-Express Etherent NICs, Single Network Port, 10GBase-SR, Optics Fiber Transceiver**

## 4.2 Experimental Evaluation

We prototyped an A-TFN system as shown in Fig. 9A. A sender connects to a receiver via two physical back-to-back 10Gbps links. The sender and receiver are the same computer systems as specified in Section 2.4. The 10Gbps links are driven by Myricom 10Gbps Ethernet NICs. In both the sender and the receiver, the two Myricom 10Gbps NICs are aggregated into a single logical bonded interface with the Linux bonding driver. In the sender, the bonding driver is modified with A-TFN mechanisms and each 10Gbps link is deemed an A-TFN receive queue. In the receiver, each slave NIC (receive queue) is pinned to a specific core. In addition, the receiver's OS is modified to support the A-TFN mechanisms. For an outgoing transport-layer packet, the OS records a processing core ID in the "transmit descriptor" and passes it to "A-TFN." Here, we make use of four reserved bits in the TCP header as the "transmit descriptor" to communicate the core ID. When the sender receives a "transmit descriptor," it extracts the passed Core ID and updates the corresponding flow entry in the Flow-to-Core table. Unless otherwise specified, $T_{timer}$ is set to 0.1 millisecond. The Flow-to-Core table is upped limited to 10, 000 entries.

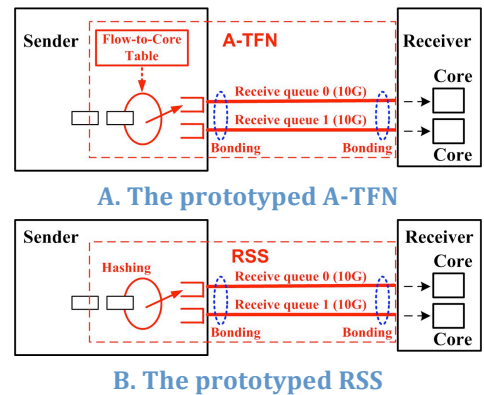

**A. The prototyped A-TFN**

**B. The prototyped RSS**

**Fig. 9 Experiment Systems**

*iperf –c receiver –P n –t 100 –p 5001 &*
*iperf –c receiver –P n  -t 100 -p 6001 &*

**Listing 1 Sender Experiment Scripts**

Similarly, we implemented a two-receive queue RSS NIC, as shown in Fig. 9B. In both the sender and the receiver, the two Myricom 10Gbps NICs are aggregated into a single logical bonded interface with the bonding driver. In the sender, the bonding driver is modified with RSS mechanisms, and each 10Gbps link is treated as an RSS receive queue. Unless otherwise specified, the hashing is based on the combination of *{src_addr, dst_addr, src_port,*

| | Experiments | Receive Queues Config. | Iperf Config. |
|---|---|---|---|
| **Performance Experiment** | Exp. 1 | Receive queue 0 @ Core 0 | "iperf –s –p 5001" @ Core {0} |
| | | Receive queue 1 @ Core 1 | "iperf –s –p 6001" @ Core {1} |
| | Exp. 2 | Receive queue 0 @ Core 0 | "iperf –s –p 5001" @ Core {0} |
| | | Receive queue 1 @ Core 2 | "iperf –s –p 6001" @ Core {2} |
| **Reordering Experiment** | Exp. 3 | Receive queue 0 @ Core 0 | "iperf –s –p 5001" @ Core {0, 1} |
| | | Receive queue 1 @ Core 1 | "iperf –s –p 6001" @ Core {0, 1} |
| | Exp.4 | Receive queue 0 @ Core 0 | "iperf –s –p 5001" @ Core {0, 2} |
| | | Receive queue 1 @ Core 2 | "iperf –s –p 6001" @ Core {0, 2} |

Table 2 Experiment Configurations in the Receiver

| $2n$ | WAITTIME-TOTAL (Nanoseconds) | | | BUS_HITM_DRV (X10000) | | |
|---|---|---|---|---|---|---|
| | RSS | A-TFN-1 | A-TFN-6 | RSS | A-TFN-1 | A-TFN-6 |
| 40 | 703819±1199 | 87371±1818 | 2913±137 | 12710±16 | 7081±31 | 6716±191 |
| 200 | 663459±27854 | 257794±10544 | 6772±330 | 11933±18 | 7691±38 | 5937±181 |
| 1000 | 738267±242107 | 472195±240886 | 68778±12502 | 7001±513 | 4620±598 | 3150±1087 |
| 2000 | 1673063±406510 | 974218±219541 | 511629±191570 | 5452±799 | 5432±106 | 3713±812 |

Table 3 Experiment 1 Results (WAITIME-TOTAL & BUS_HITM_DRV)

*dst_port}* for each incoming packet. In the receiver, each slave NIC (receive queue) is pinned to a specific core.

We ran data transmission experiments with iperf using the test system shown in Fig. 9. The experiments relied on the following: (1) Iperf is a multi-threaded network application. With multiple parallel TCP data streams, a dedicated child thread is spawned and assigned to handle each stream in the system. (2) When iperf is pinned to a specific core, its child threads are also pinned to that core. In our experiments, iperf sends from the sender to the receiver with *n* parallel TCP streams for 100 seconds, to ports 5001 and 6001, respectively. Therefore, totally *2n* parallel TCP streams are transmitting in each experiment. The number *n* was varied across experiments. The experiment scripts for the sender appear in Listing 1. In the receiver, the receive queues and iperf are pinned to different cores to simulate a two-core system. The experimental configurations are listed in Table 2.

In our emulated system, we measure the Flow-to-Core Table's search time. The search time to access the first item in a collision-resolving linked list takes around 260 *ns*, which includes the hashing and locking overheads. For each next item in the list, it takes approximately an extra 150 *ns*. Therefore, the longest search in our system takes $260 + 150 * (MaxListSize - 1)$ *ns*. For a 10Gbps NIC, the time budget to process a 1500byte packet is around 1200 *ns*. To evaluate *MaxListSize*'s effect on A-TFN's performance, we set *MaxListSize* to 1 and 6, respectively. Correspondingly, A-TFN is termed as A-TFN-1 and A-TFN-6.

*a) Performance Experiments*

Experiments 1 and 2 simulated the network conditions that a single application must handle multiple concurrent TCP connections. In both experiments, TCP streams of a specific port (5001 or 6001) were pinned to a particular core. Given the same experimental conditions, we compared the results with A-TFN to those with RSS. The metrics of interest were: (1) Throughput; (2) WAITTIME-TOTAL; and (3) BUS_HITM_DRV. (The number of events between samples was 10000.) Consistent results were obtained across repeated runs. All results presented are shown with a 95% confidence interval.

| $2n$ | Throughput (Gbps) | | |
|---|---|---|---|
| | RSS | A-TFN-1 | A-TFN-6 |
| 40 | 14.65 ± 0.04 | 14.77 ± 0.01 | 14.82 ± 0.08 |
| 200 | 13.39 ± 0.04 | 13.51 ± 0.02 | 13.62 ± 0.04 |
| 1000 | 5.93 ± 0.64 | 6.05 ± 0.87 | 6.15 ± 0.85 |
| 2000 | 5.2 ± 0.54 | 5.44 ± 0.93 | 6.06 ± 0.79 |

Table 4 Experiment 1 Results (Throughput)

As analyzed in previous sections, when a single network application handles multiple concurrent TCP connections, the hashing function of the RSS-enabled NIC will evenly and statistically distribute the connections across the cores. Since the application can only run on a single core at any given moment, some connections get steered to cores other than the one on which the application runs. As a result, TCP processing will alternate between different cores. This fact might even lead to contention for shared resources between interrupt and application threads when they concurrently process packets of the same flows. Under such circumstances, overall system efficiency could be severely degraded. The experimental results in Tables 3, 4, and Fig. 10 confirm these points. To save space, we put experiment 1's results in Table 3 and 4. In experiment 2, Core 0 and 2 reside in two separate physical processors. The results can better show the RSS limitation. We present them in Fig. 10.

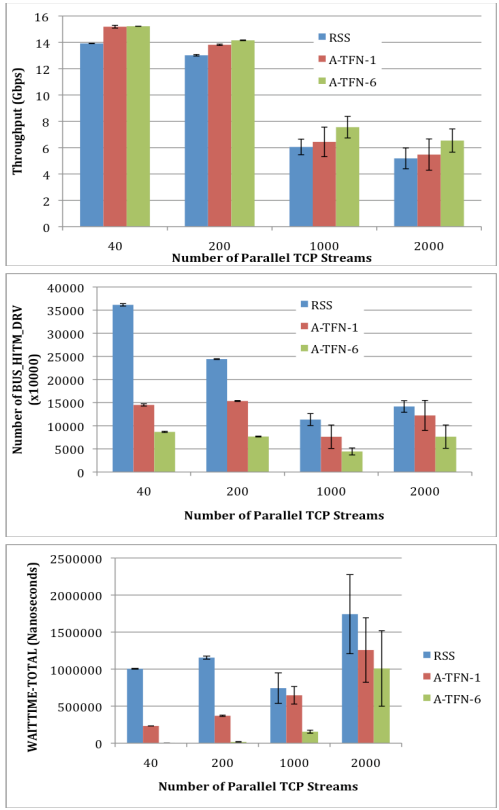

Fig. 10 Experiment 2 Results

Experiments 1 & 2 clearly show that: (1) A-TFN can effectively improve the network throughput. For example, with $2n=1000$ in experiment 2, A-TFN-6 markedly increased the TCP throughput by more than 20%. (2) A-TFN can significantly reduce lock contention in the parallel network stacks. The total time spent waiting to acquire various kernel locks was greatly decreased by more than 98% for A-TFN-6 with $2n=40$. (3) A-TFN can substantially reduce system synchronization overhead. Experimental data clearly confirm the effectiveness of A-TFN in improving network throughput and enhancing system efficiency. This is because the design of A-TFN tries to steer incoming network traffic to the cores, where they will be protocol-processed and consumed by the applications. As a result, TCP processing does not alternate between different cores, and contention involving shared resources between interrupt and application threads will not occur. Also, costly MESI operations can be greatly reduced. The experiment results further show that A-TFN is more effective in experiment 2 than in experiment 1. This because in experiment 1, core 0 and 1 reside in the same physical processor; in experiment 2, core 0 and 2 resides in different physical processors. Core-to-core synchronization is more expensive when the contending cores exist within different physical processors. Therefore, A-TFN is more effective in NUMA systems.

For the Flow-to-Core table, when a specific hash's collision-resolving lined list reaches *MaxListSize*, later arrived flows for that hash will not be entered into the table and their packets are delivered in the same way as does RSS. It can be seen that with $2n=40$, A-TFN-1's results (especially for throughputs) are very close to those of A-TFN-6's; with $2n=2000$, A-TFN-1 behaves closer as does RSS. We record the percentage of flows that are entered into the Flow-to-Core table when *n* is varied (Table 5). It clearly shows that when *n* increases, the percentage of flows that are entered into the table is decreasing; and the effects on A-TFN-1 is much more than on A-TFN-6. With $2n=2000$, A-TFN-1 has only a 12.7% of flows entered into the Flow-to-Core table. The reason that the ratio is so low is because all the flows share a single pair of IP addresses, they are not hashed across the table. As a result, more traffic would be delivered in the same way as RSS does. From the hardware implementation's perspective, A-TFN-1's Flow-to-Core table is much easier to implement. But its performance is not satisfactory when the number of TCP streams increase. There will be a tradeoff in hardware complexity and A-TFN effectiveness. It is anticipated that with *n* further increased, A-TFN-6 would have more traffic delivered in the way as RSS does; its effectiveness would start to decrease. Normally, a high-end web server would handle a few thousand concurrent TCP streams. For our two-core A-TFN emulated system, 2000 streams is quite a high number. Since the tread is already very clear, we don't further increase *n*.

| *2n* | A-TFN-6 | A-TFN-1 |
|---|---|---|
| 40 | 100% ± 0 | 88% ± 1.6% |
| 200 | 100% ± 0 | 71% ± 2.9% |
| 1000 | 95.7% ± 1.1% | 24.5% ± 0.1% |
| 2000 | 71.7% ± 0.2% | 12.7% ± 0% |

Table 5 Flows @ Flow-to-Core Table Percentage

With RSS technologies, the worst cases occur when soft partition technologies, like CPUSET, are applied in the networking environments. This can easily lead to the undesirable situation in which network applications are soft-partitioned on cores other than those to which their network interrupts are directed. Also, an OS scheduler prioritizes load balancing (or power saving) over process-to-interrupt affinity. In these environments, network applications may also be scheduled on cores other than those where their corresponding network interrupts are directed. We ran experiments in these environments. Conclusions similar to those above can be drawn, but due to space limitations, those results are not discussed here.

*b) Reordering Experiments*

A-TFN uses a special flow entry updating mechanism to guarantee in-order packet delivery. Experiments 3 and 4 were designed to evaluate whether this mechanism actually works. In both experiments, iperfs (ports 5001 and 6001) were allowed to run in both cores where the two receive queues were pinned. Linux was configured to run in *multicore peak performance* mode; the scheduler tries to use all core resources in parallel as much as possible, distributing the load equally among the cores. As a result, iperf threads may migrate across cores. The receiver was instrumented to record any out-of-order packets, and we calculated relevant packet reordering ratios. For A-TFN-6, we set $T_{timer}$ to 0 or 100 $\mu s$. The experimental results, with a 95% confidence interval, are shown in Table 6.

| $2n$ | Packet Reordering Ratio (Experiment 3) | |
|---|---|---|
| | $T_{timer} = 0\ (\mu s)$ | $T_{timer} = 100\ (\mu s)$ |
| 40 | 3.524E-07 ± 3.539E-07 | 0 |
| 200 | 7.573E-07 ± 8.569E-07 | 0 |
| 1000 | 1.252E-04 ± 1.015E-04 | 0 |
| 2000 | 2.076E-04 ± 7.200E-05 | 0 |
| $2n$ | Packet Reordering Ratio (Experiment 4) | |
| | $T_{timer} = 0\ (\mu s)$ | $T_{timer} = 100$ |
| 40 | 5.110E-07 ± 6.809E-07 | 0 |
| 200 | 6.278E-06 ± 8.553E-06 | 0 |
| 1000 | 3.639E-05 ± 2.754E-05 | 0 |
| 2000 | 2.174E-04 ± 8.515E-05 | 0 |

**Table 6 Reordering Experiments**

When $T_{timer}$ is 0, incoming packets of a flow in the transition state are immediately delivered, instead of being added to the tail of "***Packets in Transition.***" As discussed in Section 3.3, this might lead to packet reordering. The results in Table 6 reflect this fact. The reason why the packet reordering ratio was so small is because (1) the scheduler tried to the extent possible to schedule a process onto the same core on which it was previously running, so that process migration was not frequent; and (2) when a flow entry was in the transition state, its packets might not arrive during this period. This shows, from another perspective, that very few packets are held as "***Packets in Transition***," where they would experience extra delay. When $T_{timer}$ is 100 $\mu s$, no out-of-order packets are recorded. This shows that A-TFN can effectively guarantee in-order packet delivery.

## 5. Related Works

Over the years, research on affinity in network processing has been extensive. Salehi et al. [4] studied the effectiveness of affinity-based scheduling in multiprocessor network protocol processing using both packet-level and connection–level parallelization approaches. But since these approaches worked in the user space, they did not consider either system or implementation costs. In [5] and [6], A. Foong et al. experimented with affinitizing processes/threads, as well as interrupts from NICs, to specific processors in an SMP system. Experimental results suggested that processor affinity in network processing contexts can significantly improve overall performance. In [7], J. Hye-Churn et al. studied the problem of multi-core aware processor affinity for TCP/IP over multiple network interfaces, using a software-only approach. Their research topics are similar to us.

Other researchers have adopted a hard partition approach [24][25][26]. In multiprocessor environments, a subset of the processor is dedicated to network processing; the remaining processors perform only application-relevant computations. Applications interact with network processing using synchronous or asynchronous I/O interfaces. The limitation of this approach is that the OS architecture requires significant changes.

The NIC technologies, such as Intel's vmdq [27] or the PCI-SIG's SR-IOV [28], also provide data steering capabilities for the NICs. But they are I/O virtualization technologies targeting at virtual machines in the virtualized environment, with different research issues.

## 6. Conclusion and Discussion

Existing RSS-enabled NICs cannot automatically steer incoming network data to the core on which its application process resides. It causes various negative impacts. We propose an A-TFN mechanism to remedy this limitation. In the paper, we discuss two A-TFN design options. Option 1 is to minimize changes in the OS and focuses instead on identifying the minimal set of mechanisms to add to the NIC. Clearly, this design adds complexity and cost to the NIC. On the other end of the design space, it could be let the OS update the flow-to-core table directly without changing anything in the NIC hardware (option 2). Conceptually, this approach could be fairly straightforward to implement. However, it might add significant extra communication overheads between the OS and the NIC, especially when the Flow-to-Core table gets large. Due to space limitation, this paper is mainly focused on the first design option. The new NIC is emulated in software and it shows that the solution is effective and practical to remedy RSS's limitation. In our future work, we will explore the second design option.